\documentclass[12pt]{article}

\usepackage[totalwidth=475truept,totalheight=615truept]{geometry}
\usepackage{amsfonts,latexsym,amssymb,amsmath,graphicx,accents,eucal,slashed,subfigure}
\usepackage{dsfont}
\usepackage[T1]{fontenc}
\usepackage[hidelinks]{hyperref}

\global\arraycolsep=1truept

\numberwithin{equation}{section}

\begin{document}

\null

\bigskip \phantom{C}

\vskip 1.5truecm

\begin{center}
{\huge \textbf{On The Quantum Field Theory}}

\vskip.4truecm

{\huge \textbf{Of The Gravitational Interactions}}

\vskip1truecm

\textsl{Damiano Anselmi}

\vskip .2truecm

\textit{Dipartimento di Fisica ``Enrico Fermi'', Universit\`{a} di Pisa, }

\textit{Largo B. Pontecorvo 3, 56127 Pisa, Italy}

\textit{and INFN, Sezione di Pisa,}

\textit{Largo B. Pontecorvo 3, 56127 Pisa, Italy}

damiano.anselmi@unipi.it

\vskip1truecm

\textbf{Abstract}
\end{center}

We study the main options for a unitary and renormalizable, local quantum
field theory of the gravitational interactions. The first model is a
Lee-Wick superrenormalizable higher-derivative gravity, formulated as a
nonanalytically Wick rotated Euclidean theory. We show that, under certain
conditions, the $S$ matrix is unitary when the cosmological constant
vanishes. The model is the simplest of its class. However, infinitely many
similar options are allowed, which raises the issue of uniqueness. To deal
with this problem, we propose a new quantization prescription, by doubling
the unphysical poles of the higher-derivative propagators and turning them
into Lee-Wick poles. The Lagrangian of the simplest theory of quantum
gravity based on this idea is the linear combination of $R$, $R_{\mu \nu
}R^{\mu \nu }$, $R^{2}$ and the cosmological term. Only the graviton
propagates in the cutting equations and, when the cosmological constant
vanishes, the $S$ matrix is unitary. The theory satisfies the locality of
counterterms and is renormalizable by power counting. It is unique in the
sense that it is the only one with a dimensionless gauge coupling.

\vfill\eject

\section{Introduction}

\label{s0}

\setcounter{equation}{0}

The problem of quantum gravity is the compatibility between
renormalizability and unitarity. It is well known that the Hilbert-Einstein
action is not renormalizable by power counting \cite{thooftveltman}.
However, if we include the infinitely many counterterms it generates,
multiplied by independent couplings, it is perturbatively unitary \cite%
{unitarityc}. An option to improve the ultraviolet behavior of the loop
integrals is to add quadratic terms with higher derivatives. It is then
possible to build higher-derivative theories of quantum gravity that are
renormalizable with finitely many couplings \cite{stelle}. However, such
theories are not unitary, at least if they are formulated in the usual ways.

Higher-derivative theories must be formulated properly, because they are
less trivial than one would naively expect. For example, if they are defined
directly in Minkowski spacetime, i.e. by integrating the loop energies along
the real axis of the complex energy plane, they generate nonlocal,
non-Hermitian divergences when the free propagators have complex poles \cite%
{ugo}, which makes them unacceptable from the mathematical point of view. On
the other hand, the Wick rotation from Euclidean space is obstructed when
the free propagators have poles in the first or third quadrants of the
complex energy plane. The obstruction can actually be overcome by a
nonanalytic procedure, which leads to a new formulation \cite{LWformulation}
of an interesting subclass of higher-derivative theories, the Lee-Wick (LW)
models \cite{LW,LWQED}.

Viewed as nonanalytically Wick rotated Euclidean theories, such models are
perturbatively unitary \cite{LWunitarity}. Moreover, the new formulation is
intrinsically equipped with all that is needed to define the physical
amplitudes properly, with no need of \textit{ad hoc} prescriptions. The
complex energy hyperplane is divided into disjoint regions $\mathcal{A}_{i}$
of analyticity, which can be connected to one another by a well defined, but
nonanalytic procedure. It is necessary to work in suitable subsets $\mathcal{%
O}_{i}$ of the regions $\mathcal{A}_{i}$, in a generic Lorentz frame, and
analytically continue the results from $\mathcal{O}_{i}$ to $\mathcal{A}_{i}$
at the end. Finally, the nonanalytic behaviors of the physical amplitudes
suggest ways that may facilitate the experimental measurements of the key
parameters of the models.

Old formulations of the Lee-Wick models were based on \textit{ad hoc}
prescriptions, the best known one being the \textit{CLOP\ prescription} of
ref. \cite{CLOP}\footnote{%
See \cite{grinstein} for explicit calculations in this approach.}. Often,
such approaches are unambiguous in some loop diagrams, but ambiguous in
others, and do not admit a clear formulation at the Lagrangian level. In
ref. \cite{LWformulation} it has been shown that they may give ambiguous
results already at one loop.

In this paper, we investigate the main options for quantum gravity that are
offered by the nonanalytic Wick rotation of Euclidean higher-derivative
theories, combined with extra tools that we introduce anew. We begin with
the superrenormalizable Lee-Wick models, which are unitary when the
cosmological constant vanishes. We investigate the simplest representative
of this class of models in detail and show that in various cases a vanishing
cosmological constant is consistent with the renormalization group, before
and after the coupling to matter. However, the theories with similar
properties are infinitely many, which raises the issue of uniqueness. A
principle of maximum simplicity could be used to single out the model
studied here, but the principle itself would have to be justified in its
turn. For this reason, it is worth to move further on, in the search for a
unique theory of quantum gravity. We identify a candidate in a model whose
Lagrangian contains the Hilbert-Einstein term, the cosmological term, $%
R_{\mu \nu }R^{\mu \nu }$ and $R^{2}$. We formulate it by means of an
additional trick, which consists of doubling the ghost poles of the free
propagators and treating the doubled versions as Lee-Wick poles. The
perturbative unitarity of the model then follows from the one of its
Lee-Wick parent theory. The uniqueness of this options relies on the fact
that it is the only one whose gauge coupling is dimensionless (according to
the power counting of the high-energy limit).

The LW\ models have been studied in QED \cite{LWQED}, the standard model 
\cite{LWstandardM} and grand unified theories \cite{LWunification}, besides
quantum gravity \cite{LWgrav1,LWgrav2}. Although the CLOP or other \textit{%
ad hoc} prescriptions have been advocated in such investigations, some
conclusions may survive once those prescriptions are removed in favor of the
formulation of ref. \cite{LWformulation}.

We recall other options to make sense of quantum gravity that can be found
in the literature. A well known idea is asymptotic safety \cite{wein}. If
there exists an ultraviolet interacting fixed point with a finite
dimensional critical surface, then it is possible to reduce the free
parameters of quantum gravity to a finite number, by demanding that the
theory lie on the critical surface at high energies. The weakness of this
approach is that it is nonperturbative. Nevertheless, truncations and
consistency checks can provide evidence that ultraviolet fixed points may
indeed exist and have good critical surfaces \cite{reuter}.

Nonlocal theories of quantum gravity have also been explored \cite%
{kuzmin,mazumdar}. Some theories of this class are claimed to have a simple,
local renormalization \cite{kuzmin}. This may be true if they are defined in
Euclidean space, but the results of \cite{ugo} suggest that if they are
defined in Minkowski spacetime, they generate nonlocal divergences that
cannot be removed by any standard procedures. On the other hand, it is hard
to Wick rotate such nonlocal Euclidean theories, because their free
propagators contain nonpolynomial functions that have extremely involved
behaviors at infinity in the first and third quadrants of the complex energy
plane. Finally, the usual proofs of perturbative unitarity \cite{cutkosky}
do not extend to nonlocal theories straightforwardly \cite{unitarityc}.

Other possibilities to make sense of higher-derivative theories have been
explored. One is that the unphysical degrees of freedom, even if present,
might be unobservable if the renormalization group keeps their masses always
above the running energy \cite{narain}. Another possibility is that the
unphysical degrees of freedom might be a blunder due to the expansion around
the wrong vacuum.

We also recall that it is possible to treat quantum gravity as a low energy
effective field theory with infinitely many couplings. In principle, this
approach can even lead to physical predictions beyond the low-energy regime,
if we identify physical quantities that just depend on a finite subset of
parameters. For example, organizing the Lagrangian in a convenient way \cite%
{newQG}, it can be proved that the Friedmann-Lema\^{\i}tre-Robertson-Walker
(FLRW) metrics are exact solutions of the complete field equations in
arbitrary dimensions with a homogeneous and isotropic matter distribution
(after a perturbative field redefinition of the metric tensor).

The paper is organized as follows. In section \ref{QG} we study the simplest
superrenormalizable model of quantum gravity and work out the conditions
under which it is unitary. In section \ref{matt} we extend the analysis to
the coupling to matter. In section \ref{uniqueness} we address the
uniqueness problem. In section \ref{faked} we introduce the concept of fake
degree of freedom. By turning ghosts into fakes, in section \ref{qg} we
build the unique model of quantum gravity that has a dimensionless gauge
coupling and show that it is unitary up to \textquotedblleft
anomalous\textquotedblright\ effects due to the cosmological constant.
Section \ref{concl} contains our conclusions.

\section{Superrenormalizable quantum gravity}

\setcounter{equation}{0}\label{QG}

The first option that we consider is a superrenormalizable higher-derivative
gravity, formulated by nonanalytically Wick rotating its Euclidean version.
We focus on the simplest representative of this class. Up to total
derivatives, its most general Lagrangian $\mathcal{L}_{\text{QG}}$ is given
by%
\begin{eqnarray}
-2\kappa ^{2}\mu ^{\varepsilon }\frac{\mathcal{L}_{\text{QG}}}{\sqrt{-g}}
&=&2\lambda _{C}M^{2}+\zeta R-\frac{\gamma }{M^{2}}R_{\mu \nu }R^{\mu \nu }+%
\frac{1}{2M^{2}}(\gamma -\eta )R^{2}  \notag \\
&&-\frac{1}{M^{4}}(D_{\rho }R_{\mu \nu })(D^{\rho }R^{\mu \nu })+\frac{1}{%
2M^{4}}(1-\xi )(D_{\rho }R)(D^{\rho }R)  \notag \\
&&+\frac{1}{M^{4}}\left( \alpha _{1}R_{\mu \nu }R^{\mu \rho }R_{\rho }^{\nu
}+\alpha _{2}RR_{\mu \nu }R^{\mu \nu }+\alpha _{3}R^{3}+\alpha _{4}RR_{\mu
\nu \rho \sigma }R^{\mu \nu \rho \sigma }\right.  \notag \\
&&\qquad \qquad \qquad \qquad \qquad +\left. \alpha _{5}R_{\mu \nu \rho
\sigma }R^{\mu \rho }R^{\nu \sigma }+\alpha _{6}R_{\mu \nu \rho \sigma
}R^{\rho \sigma \alpha \beta }R_{\alpha \beta }^{\mu \nu }\right) ,
\label{LQG}
\end{eqnarray}%
where $\lambda _{C}$, $\zeta $, $\gamma $, $\eta $, $\xi $, $\alpha
_{1},\cdots ,\alpha _{6}$ are dimensionless constants, $\kappa $ has
dimension $-1$ in units of mass and $M$ is the Lee-Wick mass scale. The last
two lines contain a convenient basis for the six independent scalars that
can be built with three Riemann tensors.

We expand the metric tensor $g_{\mu \nu }$ around the Galilean metric $\eta
_{\mu \nu }=$diag$(1,-1,-1,-1)$ by writing%
\begin{equation*}
g_{\mu \nu }=\eta _{\mu \nu }+2\kappa h_{\mu \nu },
\end{equation*}%
where $h_{\mu \nu }$ is the quantum fluctuation. After the expansion around
flat space, we raise and lower the indices by means of the Galilean metric.
We further define $h\equiv h_{\mu }^{\mu }$. It is convenient to choose a
gauge-fixing function that is linear in the fluctuation $h_{\mu \nu }$, such
as the De Donder function 
\begin{equation*}
\mathcal{G}_{\mu }(g)=\eta ^{\nu \rho }\partial _{\rho }g_{\mu \nu }-\frac{1%
}{2}\eta ^{\nu \rho }\partial _{\mu }g_{\nu \rho }.
\end{equation*}%
We complete the gauge-fixing following the steps of ref. \cite{ugo}, so as
to obtain the gauge-fixed Lagrangian%
\begin{eqnarray*}
\mathcal{L}_{\text{gf}} &=&\mathcal{L}_{\text{QG}}+\frac{1}{4\kappa ^{2}}%
\mathcal{G}^{\mu }\left( \zeta -\gamma \frac{\square }{M^{2}}+\frac{\square
^{2}}{M^{4}}\right) \mathcal{G}_{\mu } \\
&&+\bar{C}^{\mu }\left( \zeta -\gamma \frac{\square }{M^{2}}+\frac{\square
^{2}}{M^{4}}\right) \left[ \square C_{\mu }-(2\delta _{\mu }^{\rho }\eta
^{\nu \sigma }\partial _{\nu }-\eta ^{\rho \sigma }\partial _{\mu })\Gamma
_{\rho \sigma }^{\alpha }C_{\alpha }\right] ,
\end{eqnarray*}%
where $\square =\eta ^{\mu \nu }\partial _{\mu }\partial _{\nu }$ is the
flat-space D'Alembertian.

We begin by studying the renormalization of the theory and then discuss the
conditions under which it is perturbatively unitary.

\subsection{Renormalization}

It is easy to see that the renormalization of a LW theory, formulated as the
nonanalytic Wick rotation of its Euclidean version, coincides with the
renormalization of its Euclidean version. Consider a Feynman diagram and
integrate the loop energies by means of the residue theorem, as usual. We
recall \cite{LWformulation,LWunitarity} that the nonanalytic behavior of the
Wick rotation is due to the pinching of LW poles and that the LW pinching
conditions equate a linear combination $p^{0}$ of the external energies to a
sum of frequencies $\omega _{i}(\mathbf{k},\mathbf{p})$, where $\mathbf{k}$
denotes the loop space momenta. Now, the ultraviolet divergences are studied
by keeping the external momenta $p$ fixed and letting $\mathbf{k}$ tend to
infinity. In such a limit the LW pinching conditions have no solutions,
because the frequencies $\omega _{i}$ grow linearly, but their sum is fixed.
For this reason, the LW pinching does not affect the renormalization of the
theory, which allows us to study the counterterms of $\mathcal{L}_{\text{QG}%
} $ with the usual techniques.

By power counting, the counterterms have at most dimension four. Using the
dimensional regularization, we organize them as%
\begin{equation}
\frac{\mathcal{L}_{\text{count}}}{\sqrt{-g}}=\frac{1}{(4\pi )^{2}\varepsilon 
}\left[ 2a_{C}M^{4}+a_{\zeta }M^{2}R-a_{\gamma }R_{\mu \nu }R^{\mu \nu }+%
\frac{1}{2}(a_{\gamma }-a_{\eta })R^{2}\right] ,  \label{lcount}
\end{equation}%
where $\varepsilon =4-D$, $D$ being the continued spacetime dimension. It is
easy to see that the counterterms proportional to the cosmological constant
are present up to three loops, those proportional to the Hilbert-Einstein
term are present up to two loops and the counterterms proportional to $%
R_{\mu \nu }R^{\mu \nu }$ and $R^{2}$ are just present at one loop. The
parameters $\xi $ and $\alpha _{i}$ do not run.

It is convenient to introduce the \textquotedblleft fine structure constant
of quantum gravity\textquotedblright 
\begin{equation*}
\alpha _{\text{QG}}=\frac{\kappa ^{2}M^{2}}{4\pi }.
\end{equation*}%
To minimize the number of $\pi $s in the formulas below, we also introduce
the constant%
\begin{equation*}
\bar{\alpha}=\frac{\kappa ^{2}M^{2}}{(4\pi )^{2}}=\frac{\alpha _{\text{QG}}}{%
4\pi }.
\end{equation*}

The structure of the coefficients that appear in formula (\ref{lcount}) is%
\begin{equation*}
a_{C}=a_{C}^{(1)}+u\gamma \bar{\alpha}+v\eta \bar{\alpha}+w\bar{\alpha}%
^{2},\qquad a_{\zeta }=a_{\zeta }^{(1)}+z\bar{\alpha}\qquad a_{\gamma
}=a_{\gamma }^{(1)},\qquad a_{\eta }=a_{\eta }^{(1)},
\end{equation*}%
where the superscript \textquotedblleft (1)\textquotedblright\ denotes the
one-loop values, while%
\begin{equation*}
u=u_{1}+\frac{u_{2}}{\varepsilon },\qquad v=v_{1}+\frac{v_{2}}{\varepsilon }%
,\qquad w=w_{1}+\frac{w_{2}}{\varepsilon }+\frac{w_{3}}{\varepsilon ^{2}}%
,\qquad z=z_{1}+\frac{z_{2}}{\varepsilon },
\end{equation*}%
and the coefficients $u_{i}$, $v_{i}$, $w_{i}$, $z_{i}$, are functions of
the parameters $\alpha _{i}$ and $\xi $. As usual, the renormalization group
relates the coefficients $u_{2}$, $v_{2}$, $w_{2}$, $w_{3}$ and $z_{2}$ of
the double and triple $\varepsilon $ poles to the coefficients of the simple
poles $u_{1}$, $v_{1}$, $w_{1}$ and $z_{1}$.

The bare parameters read%
\begin{eqnarray*}
\lambda _{C\text{B}} &=&\lambda _{C}-\frac{2}{\varepsilon }a_{C}\bar{\alpha}%
,\qquad \zeta _{\text{B}}=\zeta -\frac{2}{\varepsilon }a_{\zeta }\bar{\alpha}%
,\qquad \gamma _{\text{B}}=\gamma -\frac{2}{\varepsilon }a_{\gamma }\bar{%
\alpha}, \\
\eta _{\text{B}} &=&\eta -\frac{2}{\varepsilon }a_{\eta }\bar{\alpha},\qquad 
\bar{\alpha}_{\text{B}}=\bar{\alpha}\mu ^{\varepsilon }.
\end{eqnarray*}%
From these expressions, we find the beta functions%
\begin{eqnarray*}
\beta _{\zeta } &=&-2a_{\zeta }^{(1)}\bar{\alpha}-4z_{1}\bar{\alpha}%
^{2},\qquad \beta _{\gamma }=-2a_{\gamma }\bar{\alpha},\qquad \beta _{\eta
}=-2a_{\eta }\bar{\alpha}, \\
\beta _{C} &=&-2a_{C}^{(1)}\bar{\alpha}-2(2u_{1}\gamma +2v_{1}\eta +3w_{1}%
\bar{\alpha})\bar{\alpha}^{2},\qquad \beta _{\text{QG}}=0,
\end{eqnarray*}%
where $\beta _{C}$ is the beta function of $\lambda _{C}$ and $\beta _{\text{%
QG}}$ is the beta function of $\alpha _{\text{QG}}$. The cancelation of the
divergences inside the beta functions gives $u_{2}$, $v_{2}$, $w_{2}$, $%
w_{3} $ and $z_{2}$.

We have computed the one-loop counterterms in the most general case.
However, due to their involved expressions, we just report them in a
simplified case that is enough for our purposes\footnote{%
The full one-loop beta functions can be downloaded in various formats from
the website \href{http://renormalization.com}{Renormalization} at the link 
\href{http://renormalization.com/Math/QG}{betaQG}.}, i.e. at $\xi =\alpha
_{6}=0$, where we find%
\begin{eqnarray*}
a_{C}^{(1)} &=&\frac{3}{4}(4\zeta -2\gamma ^{2}+2\eta \gamma -3\eta
^{2}),\qquad \qquad \qquad a_{\zeta }^{(1)}=\frac{1}{4}\gamma \tau +\frac{1}{%
2}\eta \sigma , \\
240a_{\gamma } &=&756-1080\alpha _{1}+360\alpha _{1}^{2}+480\alpha
_{2}-480\alpha _{1}\alpha _{2}-640\alpha _{2}^{2}+960\alpha _{4}-960\alpha
_{1}\alpha _{4} \\
&&-2560\alpha _{2}\alpha _{4}-5440\alpha _{4}^{2}+1940\alpha _{5}-1080\alpha
_{1}\alpha _{5}-1280\alpha _{2}\alpha _{5}-3040\alpha _{4}\alpha
_{5}-225\alpha _{5}^{2}, \\
240a_{\eta } &=&-508-1440\alpha _{1}+1395\alpha _{1}^{2}-2400\alpha
_{2}+5160\alpha _{1}\alpha _{2}+6880\alpha _{2}^{2}-8640\alpha
_{3}+12960\alpha _{1}\alpha _{3} \\
&&+43200\alpha _{2}\alpha _{3}+77760\alpha _{3}^{2}-1920\alpha
_{4}+9600\alpha _{1}\alpha _{4}+17920\alpha _{2}\alpha _{4}+34560\alpha
_{3}\alpha _{4} \\
&&+20800\alpha _{4}^{2}+180\alpha _{5}+1170\alpha _{1}\alpha _{5}+3560\alpha
_{2}\alpha _{5}+11520\alpha _{3}\alpha _{5}+5120\alpha _{4}\alpha
_{5}+520\alpha _{5}^{2},
\end{eqnarray*}%
where%
\begin{equation*}
\tau =8+9\alpha _{1}-72\alpha _{3}+64\alpha _{4}+3\alpha _{5},\qquad \sigma
=-12+9\alpha _{1}+30\alpha _{2}+108\alpha _{3}+24\alpha _{4}+8\alpha _{5}.
\end{equation*}

\subsection{Unitarity}

To have a correct low-energy limit, we must assume $\zeta >0$. Other
important conditions on the parameters of $\mathcal{L}_{\text{QG}}$ follow
from unitarity, i.e. the very requirement that the theory is a Lee-Wick
model, which we then formulate by nonanalytically Wick rotating its
Euclidean version.

First, the extra poles of the free propagators must not be located on the
real axis, but lie symmetrically with respect to it, as in 
\begin{equation}
iS(p)=\frac{i}{p^{2}-m^{2}+i\epsilon }\,\frac{M^{4}}{(p^{2}-\mu
^{2})^{2}+M^{4}}.  \label{sp}
\end{equation}%
Second, we need to have an identically vanishing cosmological constant.
Indeed, when the cosmological constant $\lambda _{C}$ is nonvanishing, flat
space is not a solution of the field equations in the absence of matter. The
proof of unitarity cannot be carried out to the very end in that case,
because it is not known how to build conventional asymptotic states and a
consistent scattering matrix $S$ in nonflat spaces [such as (anti) de Sitter
space], although alternative approaches have been attempted \cite{adsSmatrix}%
.

We report the graviton propagator in two steps. In the relatively simple
case $\eta =\xi =0$, we have

\begin{equation}
\langle h_{\mu \nu }(p)\hspace{0.01in}h_{\rho \sigma }(-p)\rangle _{\eta
=\xi =0}^{\text{free}}=\frac{iM^{4}}{2}\frac{\eta _{\mu \rho }\eta _{\nu
\sigma }+\eta _{\mu \sigma }\eta _{\nu \rho }-\eta _{\mu \nu }\eta _{\rho
\sigma }}{P(1,\gamma ,\zeta ,2\lambda _{C})},  \label{props}
\end{equation}%
where%
\begin{equation*}
P(a,b,c,d)\equiv a(p^{2})^{3}+bM^{2}(p^{2})^{2}+cM^{4}p^{2}+dM^{6}.
\end{equation*}%
At nonvanishing $\eta $ and $\xi $ the free propagator reads%
\begin{eqnarray}
\langle h_{\mu \nu }(p)\hspace{0.01in}h_{\rho \sigma }(-p)\rangle ^{\text{%
free}} &=&\langle h_{\mu \nu }(p)\hspace{0.01in}h_{\rho \sigma }(-p)\rangle
_{\eta =\xi =0}^{\text{free}}  \notag \\
&&-\frac{iM^{4}(\eta M^{2}+\xi p^{2})}{2P(1,\gamma ,\zeta ,2\lambda _{C})}%
\frac{(p^{2}\eta _{\mu \nu }+2p_{\mu }p_{\nu })(p^{2}\eta _{\rho \sigma
}+2p_{\rho }p_{\sigma })}{P(1-3\xi ,\gamma -3\eta ,\zeta ,2\lambda _{C})}.
\label{propac}
\end{eqnarray}

As said, we must ensure that the denominators have the correct LW form at $%
\lambda _{C}=0$, as in (\ref{sp}). This happens (with $m=0$), if%
\begin{equation}
\gamma ^{2}<4\zeta ,\qquad (\gamma -3\eta )^{2}<4\zeta (1-3\xi ),\qquad
\gamma <0,\qquad \gamma <3\eta .  \label{unitbo0}
\end{equation}

We must also ensure that the renormalization group is compatible with an
identically vanishing cosmological constant. We begin by studying this issue
in the simple case $\lambda _{C}=\eta =\xi =\alpha _{6}=0$. Some unitarity
bounds (\ref{unitbo0}) become identical, so we just have%
\begin{equation}
\gamma ^{2}<4\zeta ,\qquad \gamma <0.  \label{unitbo}
\end{equation}

While the conditions $\xi =\alpha _{6}=0$ are preserved by renormalization,
because $\xi $ and $\alpha _{6}$ do not run, the conditions $\lambda
_{C}=\eta =0$ must be accompanied, for consistency with renormalization
group invariance, by%
\begin{equation}
\beta _{C}=\beta _{\eta }=0.  \label{cancel}
\end{equation}%
The condition $\beta _{\eta }=0$ is relatively easy to solve (see below) and
constrains the parameters $\alpha _{i}$. The condition $\beta _{C}=0$ gives 
\begin{equation}
3(2\zeta -\gamma ^{2})+2(2u_{1}\gamma +3w_{1}\bar{\alpha})\bar{\alpha}=0
\label{ho}
\end{equation}%
and is also easy to solve, because it just gives $\zeta $ as a function of
the other parameters.

Renormalitazion group invariance demands that the beta function of (\ref{ho}%
) be also zero, which gives%
\begin{equation}
3\gamma \left( 4a_{\gamma }-\tau \right) =8\left( 3z_{1}+u_{1}a_{\gamma
}\right) \bar{\alpha}.  \label{host}
\end{equation}%
In turn, the beta function of this relation must vanish, which implies%
\begin{equation}
\left( 4a_{\gamma }-\tau \right) a_{\gamma }=0.  \label{hol}
\end{equation}%
The beta function of this relation is identically zero, so the list of
consistency conditions stops here.

Equation (\ref{hol}) implies either $a_{\gamma }=0$ or $a_{\gamma }=-\tau /4$%
. The second possibility implies that the right-hand side of (\ref{host}) is
zero. However, this condition requires knowledge about the two-loop
renormalization of the theory, because it involves $u_{1}$ and $z_{1}$. For
this reason, we solve (\ref{hol}) by setting $a_{\gamma }=0$, to work out
solutions that just need the renormalization at one loop.

Now, $a_{\gamma }=0$ implies $\beta _{\gamma }=0$ and ensures that $\gamma $
is a number that we can choose at will. Equations (\ref{host}) then gives $%
\beta _{\zeta }=0$, which leads to $\lambda _{C}=\eta =\beta _{C}=\beta
_{\eta }=\beta _{\gamma }=\beta _{\zeta }=0$, which means that the
Lagrangian (\ref{LQG}) must be completely finite.

Summarizing, to enforce finiteness, we must solve the system of equations%
\begin{equation}
\zeta =\frac{\gamma ^{2}}{2}-\frac{1}{3}(2u_{1}\gamma +3w_{1}\bar{\alpha})%
\bar{\alpha},\qquad a_{\gamma }=a_{\eta }=0,\qquad \tau \gamma =-8z_{1}\bar{%
\alpha}.  \label{solv}
\end{equation}%
The first condition just gives $\zeta $ in terms of the other constants.
Since $\zeta $ must be nonvanishing at $\bar{\alpha}=0$, because it
multiplies the Hilbert-Einstein term, the same must be true of $\gamma $.

The easiest way to show that solutions do exist is to work them out at $\bar{%
\alpha}=0$ and check that they can be extended perturbatively to arbitrary $%
\bar{\alpha}$. Then, at $\bar{\alpha}=0$ we must take 
\begin{equation*}
\zeta =\frac{\gamma ^{2}}{2},
\end{equation*}%
which is compatible with the unitarity bound (\ref{unitbo}). At this point,
we must solve the system%
\begin{equation*}
a_{\gamma }=a_{\eta }=\tau =0,
\end{equation*}%
for the parameters $\alpha _{i}$. The condition $\tau =0$ is linear in $%
\alpha _{i}$ and can be easily solved for one of such parameters. Inserting
the solution into the conditions $a_{\gamma }=a_{\eta }=0$, we get two
coupled quadratic equations in five unknowns. Acceptable solutions are easy
to find algebraically, after setting three parameters $\alpha _{i}$ to zero.
For example, we set $\alpha _{4}=\alpha _{5}=\alpha _{6}=0$ and solve $\tau
=0$ for $\alpha _{3}$. So doing, we obtain a system of two quadratic
equations $e_{1}(\alpha _{1},\alpha _{2})=e_{2}(\alpha _{1},\alpha _{2})=0$
in the two unknowns $\alpha _{1},\alpha _{2}$. The four real solutions at $%
\alpha _{4}=\alpha _{5}=\alpha _{6}=0$ are%
\begin{eqnarray}
(\alpha _{1},\alpha _{2}) &=&(4.51163..,-3.91524...),\qquad
(2.89114...,-1.93684...),  \notag \\
&&(0.800169...,-0.368609...),\qquad (0.197062...,-0.679314...),  \label{unb}
\end{eqnarray}%
while $\alpha _{3}=(8+9\alpha _{1})/72$. All such solutions can be extended
perturbatively to nonvanishing $\bar{\alpha}$, because the matrices of the
derivatives $\partial e_{i}/\partial \alpha _{j}$, $i,j\leqslant 2$, are
nonsingular on (\ref{unb}).

In the end, a simple example of a consistent theory of pure quantum gravity
is the one with the Lagrangian%
\begin{eqnarray*}
\mathcal{L} &=&-\frac{\sqrt{-g}}{2\kappa ^{2}}\mu ^{-\varepsilon }\left[ 
\frac{\gamma ^{2}}{2}R-\frac{\gamma }{M^{2}}\left( R_{\mu \nu }R^{\mu \nu }-%
\frac{1}{2}R^{2}\right) -\frac{1}{M^{4}}(D_{\rho }R_{\mu \nu })(D^{\rho
}R^{\mu \nu })+\frac{1}{2M^{4}}(D_{\rho }R)(D^{\rho }R)\right. \\
&&\qquad \qquad \qquad \qquad \quad \left. +\frac{1}{M^{4}}\left( \alpha
_{1}^{\ast }R_{\mu \nu }R^{\mu \rho }R_{\rho }^{\nu }+\alpha _{2}^{\ast
}RR_{\mu \nu }R^{\mu \nu }+\frac{1}{72}(8+9\alpha _{1}^{\ast })R^{3}\right) +%
\mathcal{O}(\bar{\alpha})\right] ,
\end{eqnarray*}%
where $\gamma $ is negative but arbitrary, $\alpha _{1,2}^{\ast }$ are the
values listed in (\ref{unb}) and the corrections $\mathcal{O}(\bar{\alpha})$
are determined as explained above.

Now we repeat the analysis relaxing the simplifying condition $\eta =0$,
i.e. we just assume $\xi =\alpha _{6}=0$ and study the conditions under
which the cosmological constant $\lambda _{C}$ vanishes identically. The
theory is no longer finite, because the parameters $\zeta $, $\gamma $ and $%
\eta $ can run.

The renormalization group consistency conditions generated by $\lambda
_{C}=0 $ give, at $\bar{\alpha}=0$,%
\begin{eqnarray*}
\zeta &=&\frac{1}{2}\gamma ^{2}-\frac{1}{2}\eta \gamma +\frac{3}{4}\eta
^{2},\qquad \gamma =-2\eta \frac{\sigma +a_{\gamma }-3a_{\eta }}{\tau
-4a_{\gamma }+2a_{\eta }}, \\
&&a_{\gamma }(\tau -4a_{\gamma }+2a_{\eta })+2a_{\eta }(\sigma +a_{\gamma
}-3a_{\eta })=0.
\end{eqnarray*}%
An example of acceptable solution is (rounding to four decimal places) 
\begin{equation}
\zeta =43.6118\eta ^{2},\qquad \gamma =-8.7722\eta ,\qquad \alpha
_{3}=-0.0367,  \label{run}
\end{equation}%
together with $\lambda _{C}=\xi =\alpha _{1}=\alpha _{2}=\alpha _{4}=\alpha
_{5}=\alpha _{6}=0$. It is easy to check that the unitarity bounds (\ref%
{unitbo0}) are satisfied if $\eta >0$. Moreover, the solutions can be
extended to $\bar{\alpha}\neq 0$. The $\eta $ running is given by $\beta
_{\eta }=0.7182\bar{\alpha}$ and the runnings of $\zeta $ and $\gamma $
follow from their relations with $\zeta $.

Once we have a model where the cosmological constant vanishes identically,
the asymptotic states and the $S$ matrix can be defined in the usual way.
Then, the proof of perturbative unitarity can be worked out by combining the
strategy of \cite{LWunitarity} (to show that the Lee-Wick poles do not
propagate through the cuts in the cutting equations), with the strategy of
ref. \cite{unitarityc} (to show that the gauge degrees of freedom -- i.e.
those propagated by the gauge-dependent poles of $h_{\mu \nu }$ and the
Faddeev-Popov ghosts -- also do not propagate through the cuts). In
particular, we must work in a gauge that interpolates between the Coulomb
one and the one we used to derive the propagators (\ref{props}) and (\ref%
{propac}). Artificial masses $m_{g}$ for the graviton are introduced to have
control on the infrared divergences. When $m_{g}\neq 0$ the gauge degrees of
freedom drop out by approaching the Coulomb limit arbitrarily without
reaching it. After that, the limit $m_{g}\rightarrow 0$ can be safely taken
in suitable combinations of amplitudes where the infrared divergences
mutually cancel out. For details, see ref. \cite{unitarityc}.

\section{Coupling to matter}

\setcounter{equation}{0}\label{matt}

Now we generalize the analysis of the previous section to the coupling to
matter. In this context, \textquotedblleft matter\textquotedblright\ refers
to every classical field but the graviton, including gauge vectors. To begin
with, we assume that the matter fields are massless and switch off all the
matter self interactions. In other words, we take free massless scalars,
fermions and vectors, and covariantize their actions to couple them to
gravity. The matter Lagrangian $\mathcal{L}_{m}$ is given by%
\begin{equation*}
\frac{\mathcal{L}_{m}}{\sqrt{-g}}=-\frac{1}{4}F_{\mu \nu }F^{\mu \nu }+i\bar{%
\psi}e_{a}^{\mu }\gamma ^{a}D_{\mu }\psi +\frac{1}{2}(\partial _{\mu
}\varphi )g^{\mu \nu }(\partial _{\nu }\varphi )+\frac{1}{12}(1+2\varpi
)R\varphi ^{2},
\end{equation*}%
and the counterterms are \cite{hathrell,freeman}%
\begin{equation}
\frac{\mathcal{L}_{m\hspace{0.01in}\text{count}}}{\sqrt{-g}}=\frac{\mu
^{-\varepsilon }}{(4\pi )^{2}\varepsilon }\left[ -2c\left( R_{\mu \nu
}R^{\mu \nu }-\frac{1}{3}R^{2}\right) -\frac{n_{s}\varpi ^{2}}{18}R^{2}%
\right] ,  \label{conta}
\end{equation}%
where $c$ is known as \textquotedblleft central charge\textquotedblright\ in
conformal field theory, equal to 
\begin{equation*}
c=\frac{1}{120}(n_{s}+6n_{f}+12n_{v}).
\end{equation*}%
Here $n_{s}$ denotes the number of real scalar fields, $n_{f}$ is the number
of Dirac fermions plus one half the number of Weyl fermions, while $n_{v}$
is the number of vector fields. Note that the right-hand side of (\ref{conta}%
) is proportional to the square of the Weyl tensor at $\varpi =0$, the
reason being that the matter action is Weyl invariant in that case.

The counterterms (\ref{conta}) only affect the beta functions $\beta
_{\gamma }$ and $\beta _{\eta }$. The corrected beta functions are obtained
by making the replacements%
\begin{equation}
a_{\gamma }\rightarrow a_{\gamma }+2c,\qquad a_{\eta }\rightarrow a_{\eta }+%
\frac{2}{3}c+\frac{n_{s}\varpi ^{2}}{9}.  \label{repla}
\end{equation}%
In the end, we have%
\begin{eqnarray}
\beta _{\zeta } &=&-\frac{1}{2}\gamma \tau \bar{\alpha}-\eta \sigma \bar{%
\alpha}-4z_{1}\bar{\alpha}^{2},\qquad \beta _{\gamma }=-2(a_{\gamma }+2c)%
\bar{\alpha},\qquad \beta _{\eta }=-2\left( a_{\eta }+\frac{2}{3}c\right) 
\bar{\alpha}-\frac{2n_{s}\varpi ^{2}}{9}\bar{\alpha},  \notag \\
\beta _{C} &=&-\frac{3}{2}(4\zeta -2\gamma ^{2}+2\eta \gamma -3\eta ^{2})%
\bar{\alpha}-2(2u_{1}\gamma +2v_{1}\eta +3w_{1}\bar{\alpha})\bar{\alpha}%
^{2},\qquad \beta _{\text{QG}}=0.  \label{betatota}
\end{eqnarray}%
where the higher-loop contributions need not coincide with those of the pure
theory.

We inquire when we can prove perturbative unitarity again. In the simple
case $\lambda _{C}=\eta =\xi =\alpha _{6}=0$, we impose (\ref{cancel}) and
the consistency conditions that follow from renormalization group
invariance. As before, the condition $\beta _{C}=0$ can be solved
immediately, since it just gives $\zeta $ in terms of the other parameters.
The consistency conditions that make the solution renormalization group
invariant can be obtained by making the replacements (\ref{repla}) inside (%
\ref{solv}). The net result is again that the theory must be completely
finite. We must solve%
\begin{equation*}
a_{\gamma }+2c=0,\qquad a_{\eta }+\frac{2}{3}c=-\frac{n_{s}\varpi ^{2}}{9}%
,\qquad \gamma \tau =-8z_{1}\bar{\alpha}.
\end{equation*}%
Various acceptable solutions exist, with the matter content of the standard
model ($n_{s}=4$, $n_{f}=45/2$, $n_{v}=12$). An example of unitary quantum
gravity coupled to massless non-self-interacting matter is the model which
at $\bar{\alpha}=0$ has parameters%
\begin{eqnarray*}
\zeta &=&\frac{\gamma ^{2}}{2},\qquad \alpha _{1}=8\left( \alpha _{3}-\frac{1%
}{9}\right) ,\qquad \alpha _{2}=-3.3306,\qquad \alpha _{3}=0.5721, \\
\lambda _{C} &=&\eta =\varpi =\xi =\alpha _{4}=\alpha _{5}=\alpha _{6}=0,
\end{eqnarray*}%
while the corrections for $\bar{\alpha}\neq 0$ are determined with the
procedure explained before.

At very large distances, the standard model loses the QCD sector, the
massive vector bosons as well as all the other massive particles. Only the
free photon survives. If we couple it to the quantum gravity theory (\ref%
{LQG}), we obtain ($n_{s}=n_{f}=0$, $n_{v}=1$) for example the solution 
\begin{eqnarray}
\zeta &=&\frac{\gamma ^{2}}{2},\qquad \alpha _{1}=8\left( \alpha _{3}-\frac{1%
}{9}\right) ,\qquad \alpha _{2}=-3.8957,\qquad \alpha _{3}=0.6716,  \notag \\
\lambda _{C} &=&\eta =\varpi =\xi =\alpha _{4}=\alpha _{5}=\alpha _{6}=0.
\label{unitary}
\end{eqnarray}

When we add the masses and the matter self interactions, the beta functions $%
\beta _{C}$ receives nontrivial corrections. In general, setting it to zero
and imposing the consistency conditions required by renormalization group
invariance leads to a finite theory.

Call the couplings of the theory $\lambda _{i}$ and denote their beta
functions by $\beta _{i}$. Assume we want to set some function $f(\lambda )$
of the couplings to zero. Then we must also set the beta function of $%
f(\lambda )$ to zero, and the beta function of the beta function, etc. We
get the system of equations%
\begin{eqnarray}
f &=&0,\qquad \beta _{i}f_{i}=0,\qquad \beta _{j}(\beta _{ij}f_{i}+\beta
_{i}f_{ij})=0,\qquad  \notag \\
&&\beta _{k}(\beta _{jk}\beta _{ij}f_{i}+\beta _{j}\beta _{ijk}f_{i}+3\beta
_{j}\beta _{ij}f_{ik}+\beta _{j}\beta _{i}f_{ijk})=0,  \label{chain}
\end{eqnarray}%
etc., where $\beta _{ij_{1}\cdots j_{n}}\equiv \partial \beta _{i}/(\partial
\lambda _{j_{1}}\cdots \partial \lambda _{j_{n}})$ and $f_{j_{1}\cdots
j_{n}}\equiv \partial f/(\partial \lambda _{j_{1}}\cdots \partial \lambda
_{j_{n}})$. We call the conditions (\ref{chain}) the \textit{renormalization
group (RG)\ chain} generated by $f$. In our cases, the \textit{unitarity
chain} is the one starting from $f(\lambda )=\lambda _{C}$. In general, the
RG chain is made of independent equations for the parameters $\lambda _{i}$
and so fixes all of them to constant values (assuming that such values do
exist and are physically acceptable), which means that the theory is finite.
When the beta functions are particularly simple, there may be exceptions
where some parameters run after imposing the RG chain, as shown in the
example (\ref{run}).

In realistic models, the cosmological constant does not vanish identically,
so we cannot prove perturbative unitarity in a strict sense. It might be
possible to prove (a generalized notion of) perturbative unitarity in an
unconventional approach, but we do not know this for sure at present. The
other option is that unitarity is anomalous in the universe and the
cosmological constant is the measure of such an anomaly.

It is worth noting that, under some assumptions, the proof of the cutting
equations formally works even when the cosmological constant $\lambda _{C}$
is negative, because in that case the propagator acquires a sort of mass
term. Although flat space is no longer a solution of the classical field
equations, we can still expand around it, since the physics does not depend
on the expansion we choose. Consider again the simple case $\eta =\xi =0$.
Using the same gauge fixing as we used at $\lambda _{C}=0$, the graviton
propagator becomes (\ref{props}). Defining $x=p^{2}/M^{2}$,
\textquotedblleft unitarity\textquotedblright\ requires that the polynomial $%
2\lambda _{C}+\zeta x+\gamma x^{2}+x^{3}$ that appears in the denominator
have the form $(x-a)((x-b)^{2}+c)$ with $a\geqslant 0$, $b\geqslant 0$, $c>0$%
. Necessary conditions are $2\lambda _{C}=-a(b^{2}+c)<0$, as well as $\zeta
>0$, $\gamma <0$. The simplest case where such $a$, $b$, $c$ exist is when
the parameters also satisfy $\gamma \zeta =2\lambda _{C}$. Then, we have $%
a=-2\lambda _{C}/\zeta $, $b=0$, $c=\zeta $. When $\gamma \zeta \leqslant
2\lambda _{C}$, $b$ is positive and the solution continues to exist as long
as $c$ stays positive. In these cases, the derivation of the cutting
equations formally extends to the case of nonvanishing cosmological
constant. Moreover, the examples of finite theories considered so far remain
finite even when $\lambda _{C}\neq 0$, since the beta functions receive no
contributions from the cosmological constant.

\section{The problem of uniqueness}

\setcounter{equation}{0}\label{uniqueness}

The theory with Lagrangian (\ref{LQG}) is the simplest model belonging to
the class of superrenormalizable theories of quantum gravity. The other
models can be obtained from (\ref{LQG}) by adding more and more higher
derivatives and fulfilling the constraints due to unitarity and
renormalizability by power counting. The Lagrangians are 
\begin{equation*}
-2\kappa ^{2}\mu ^{\varepsilon }\frac{\mathcal{L}_{\text{QG}}^{\prime }}{%
\sqrt{-g}}=2\lambda _{C}M^{2}+\zeta R+\frac{1}{M^{2}}R_{\mu \nu
}P_{n}(\square _{c}/M^{2})R^{\mu \nu }-\frac{1}{2M^{2}}RQ_{n}(\square
_{c}/M^{2})R+V(R),
\end{equation*}%
where $\square _{c}$ denotes the covariant D'Alembertian, $P_{n}$, $Q_{n}$
are real polynomials of degree $n>1$ and $V(R)$ is a linear combination of
scalars that have dimensions ranging from 6 to $2n+4$ and are built with at
least three Riemann tensors (or their covariant derivatives).

In the extended class, it is easier to set the cosmological constant to zero
at all energies. For example, for $V(R)=0$, $n>2$ there are only one-loop
counterterms and the beta function $\beta _{C}$ of the cosmological constant
is a linear combination of the coefficients $\zeta _{1}$ and $\zeta _{2}$ of
the terms $R_{\mu \nu }\square _{c}^{n-2}R^{\mu \nu }$ and $R\square
_{c}^{n-2}R$, plus a quadratic polynomial in the coefficients $\gamma $ and $%
\eta $ of $R_{\mu \nu }\square _{c}^{n-1}R^{\mu \nu }$ and $R\square
_{c}^{n-1}R$. Moreover, $\zeta _{1}$, $\zeta _{2}$, $\gamma $ and $\eta $ do
not run. Setting $\beta _{C}=0$ gives a relation between such constants,
which can be easily solved for $\zeta _{1}$ or $\zeta _{2}$ in terms of the
other three. The RG\ chain stops immediately. If the denominators of the
propagators have the right form, the theory is also unitary. The beta
function $\beta _{\zeta }$ is linear in $\gamma $ and $\eta $. The couplings
of the quadratic terms $R_{\mu \nu }R^{\mu \nu }$ and $R^{2}$ may also have
nontrivial beta functions.

In the end, the superrenormalizable models of quantum gravity that are
unitary are infinitely many, which leads to a lack of uniqueness. The theory
(\ref{LQG}) is singled out among the others if we accept a sort of
\textquotedblleft minimum principle\textquotedblright , stating that the
right theory is just the simplest one. However, it would be better to have a
really unique answer.

A possibility would be a theory with a dimensionless gauge coupling, that is
to say a strictly renormalizable theory, which would make quantum gravity
more similar to the other gauge theories. A conventional strictly
renormalizable Lee-Wick model of quantum gravity in four dimensions does not
exist, because the propagators would not have the structure (\ref{sp}). For
this reason, we need an improved approach to the problem.

\section{Fake degrees of freedom}

\setcounter{equation}{0}\label{faked}

In this section we investigate the idea of doubling the ghost poles of the
free propagators and turn them into LW poles. An extra, fictitious LW scale $%
\mathcal{E}$ is introduced and removed at the end. This leads to a new
quantization prescription. In the next section we explore the consequences
of this idea in quantum gravity.

Start from the (massless) $\varphi ^{4}$ scalar field theory%
\begin{equation*}
\mathcal{L}=\frac{1}{2}(\partial _{\mu }\varphi )(\partial ^{\mu }\varphi )-%
\frac{\lambda }{4!}\varphi ^{4},
\end{equation*}%
in four dimensions and formulate it in Euclidean space. We write the
Euclidean propagator $1/p_{E}^{2}$ as $p_{E}^{2}/(p_{E}^{2})^{2}$, where $%
p_{E}$ is the Euclidean momentum. Then, we deform the propagator with the
help of the fictitious LW scale $\mathcal{E}$ into%
\begin{equation}
\frac{p_{E}^{2}}{(p_{E}^{2})^{2}+\mathcal{E}^{4}}  \label{pea}
\end{equation}%
and consider the limit $\mathcal{E}\rightarrow 0$. If we first let $\mathcal{%
E}$ tend to zero and then Wick rotate, we obtain the usual scalar field
theory. On the other hand, if we first Wick rotate, then let the scale $%
\mathcal{E}$ tend to zero, the $S$ matrix is identically one. Indeed, at $%
\mathcal{E}>0$ we obtain a Lee-Wick model. Formulated as a nonanalytically
Wick rotated Euclidean theory, it is perturbatively unitary and has no
physical degree of freedom.

We point out that the prescription we have just defined does not give the
principal value of $-1/p^{2}$ after the Wick rotation. Indeed, since we are
coming from the Euclidean space, the poles $p^{0}=\sqrt{\mathbf{p}^{2}\pm i%
\mathcal{E}^{2}}$ are located below the integration path, while the poles $%
p^{0}=-\sqrt{\mathbf{p}^{2}\pm i\mathcal{E}^{2}}$ are located above.
Instead, the principal value places the poles $p^{0}=\pm \sqrt{\mathbf{p}%
^{2}\pm i\mathcal{E}^{2}}$ above the integration path and the poles $%
p^{0}=\mp \sqrt{\mathbf{p}^{2}\pm i\mathcal{E}^{2}}$ below.

For this reason, our construction defines a distribution of a new type.
Flipping the overall sign, after the Wick rotation we write it as%
\begin{equation}
\lim_{\mathcal{E}\rightarrow 0}\frac{p^{2}}{[(p^{2})^{2}+\mathcal{E}^{4}]_{%
\text{LW}}}.  \label{newdistr}
\end{equation}%
The subscript \textquotedblleft LW\textquotedblright\ in the denominator is
to remind us about the positions of the poles with respect to the
integration path (the right pair of poles being placed below and the left
pair being placed above).

A good check that (\ref{newdistr}) is well defined can be made by
calculating the bubble diagram explicitly with the technique explained in
ref. \cite{LWformulation} and then take the limit $\mathcal{E}\rightarrow 0$%
. The calculation can be carried out to the very end and, after
renormalizing the ultraviolet divergence, gives (for $p$ real)%
\begin{equation}
-\frac{i}{4(4\pi )^{2}}\ln \frac{(p^{2})^{2}}{\mu ^{4}},  \label{ampla}
\end{equation}%
where we have included the combinatorial factor $1/2$. The complex energy
plane is divided into three disjoint regions. The main region $\mathcal{A}%
_{0}$ is the one that contains the imaginary axis, where the Wick rotation
is analytic. The other two regions $\mathcal{A}_{1}$ and $\mathcal{A}%
_{1}^{\prime }$ are symmetric with respect to the imaginary axis and
intersect the real axis in the half lines $p^{0}>|\mathbf{p}|$ and $p^{0}<-|%
\mathbf{p}|$. At $\mathbf{p}\neq 0$ the half $p^{0}$ plane with $\mathrm{Re}%
[p^{0}]\geqslant 0$ looks like 
\begin{equation}
\includegraphics[width=6truecm]{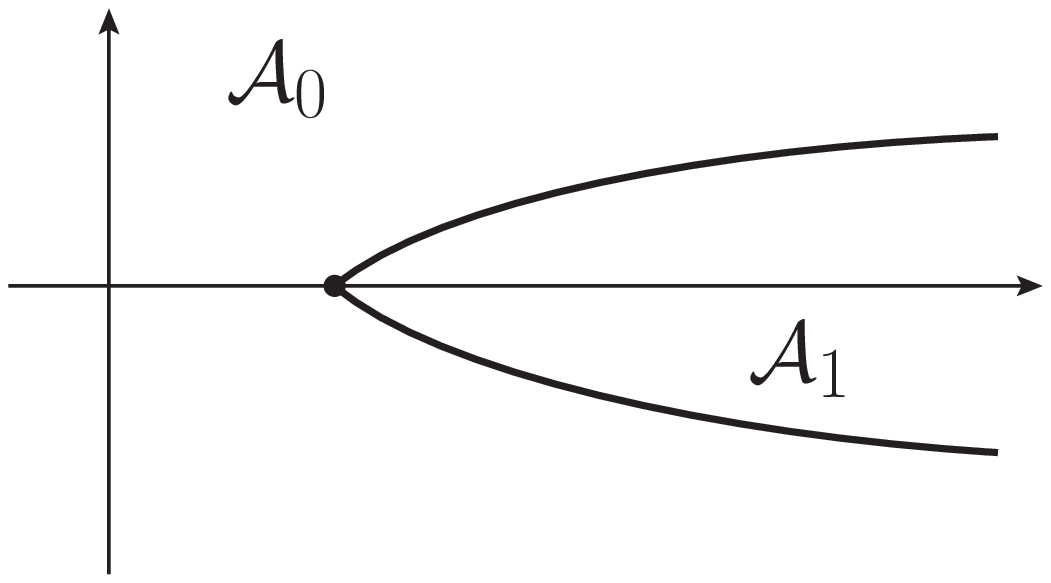}  \label{Regions}
\end{equation}%
The boundary separating the regions $\mathcal{A}_{0}$ and $\mathcal{A}_{1}$
can be deformed arbitrarily, as long as it does not intersect the real axis
anywhere but in the threshold $p^{0}=|\mathbf{p}|$. The real part of the
diagram vanishes for all real values of $p$, which confirms that the $S$
matrix is identically 1. We may say that (\ref{newdistr}) turns the degree
of freedom of the $\varphi ^{4}$ theory into a \textit{fake} degree of
freedom.

Let us compare this result with the one given by the Feynman prescription,
which is%
\begin{equation}
-\frac{i}{2(4\pi )^{2}}\ln \frac{-p^{2}-i\epsilon }{\mu ^{2}}.
\label{amplaf}
\end{equation}%
The two expressions (\ref{ampla}) and (\ref{amplaf}) coincide in the main
region $\mathcal{A}_{0}$. However, the Feynman prescription does not predict
the region $\mathcal{A}_{1}$, but maximally extends $\mathcal{A}_{0}$, so
the amplitude (\ref{amplaf}) has a discontinuity on the real axis above the
threshold $p^{0}=|\mathbf{p}|$.

With a similar procedure, we can turn the ghosts into fake degrees of
freedom. Consider the higher-derivative theory with Lagrangian%
\begin{equation*}
\mathcal{L}=\frac{1}{2}(\partial _{\mu }\varphi )\left( \zeta -\gamma \frac{%
\square }{M^{2}}\right) (\partial ^{\mu }\varphi )-\frac{\lambda }{4!}%
\varphi ^{4},
\end{equation*}%
with $\zeta >0$, $\gamma <0$. The Euclidean propagator is%
\begin{equation}
\frac{M^{2}}{p_{E}^{2}(\zeta M^{2}-\gamma p_{E}^{2})}=\frac{1}{\zeta
p_{E}^{2}}+\frac{\gamma }{\zeta (\zeta M^{2}-\gamma p_{E}^{2})}  \label{eup}
\end{equation}%
and its naive Wick rotation to Minkowski spacetime propagates a physical
massless scalar and a massive ghost. Let us deform (\ref{eup}) into%
\begin{equation*}
\frac{1}{\zeta p_{E}^{2}}+\frac{\gamma (\zeta M^{2}-\gamma p_{E}^{2})}{\zeta %
\left[ (\zeta M^{2}-\gamma p_{E}^{2})^{2}+\mathcal{E}^{4}\right] }.
\end{equation*}%
After the Wick rotation, we find (multiplying by $-i$)%
\begin{equation*}
\frac{i}{\zeta (p^{2}+i\epsilon )}-\frac{i\gamma (\zeta M^{2}+\gamma p^{2})}{%
\zeta \left[ (\zeta M^{2}+\gamma p^{2})^{2}+\mathcal{E}^{4}\right] _{\text{LW%
}}},
\end{equation*}%
which just propagates a massless particle, since the poles at $p^{2}=-(\zeta
M^{2}\pm i\mathcal{E}^{2})/\gamma $ compensate each other in the cut
propagators for every $\mathcal{E}>0$, then also for $\mathcal{E}\rightarrow
0$.

The procedure is very general and can be used to make sense of
higher-derivative quantum gravity, by turning its ghosts into fakes, as we
explain in the next section.

We have said in section \ref{QG} that the renormalization of the theory at $%
\mathcal{E}\neq 0$ coincides with the renormalization of its Euclidean
version. It is easy to show that this property survives the limit $\mathcal{E%
}\rightarrow 0$. We know that the complex energy plane is divided into
disjoint regions at $\mathcal{E}\neq 0$, the main one being the region that
contains the imaginary axis, where the Wick rotation is analytic. The other
regions are related to the main one by means of a nonanalytic procedure.
Renormalizability holds because the divergent parts of the amplitudes just
concern the main region. In the limit $\mathcal{E}\rightarrow 0$ the
intersection between the main region and the real axis is the interval $-|%
\mathbf{p}|<p^{0}<|\mathbf{p}|$.

For example, in the case of the bubble diagram, two LW pinchings have
thresholds on the real axis, their threshold being $p^{2}=2\mathcal{E}^{2}$.
The condition of pinching is%
\begin{equation*}
\pm p^{0}=\sqrt{\mathbf{k}^{2}+i\mathcal{E}^{2}}+\sqrt{(\mathbf{k-p})^{2}-i%
\mathcal{E}^{2}},
\end{equation*}%
where $p$ is the external momentum and $k$ is the loop momentum. This
condition cannot be solved for arbitrarily large loop space momentum $%
\mathbf{k}$. Therefore, the ultraviolet divergences are not affected by the
LW pinching. The conclusion also holds at $\mathcal{E}\rightarrow 0$, where
the condition becomes 
\begin{equation*}
|p^{0}|=|\mathbf{k|}+|\mathbf{k-p}|,
\end{equation*}%
which is also bounded in $|\mathbf{k|}$ when the external momentum is fixed.

We can use the new distribution (\ref{newdistr}) to build theories with
unforeseen properties. For example, consider the theory with Lagrangian%
\begin{equation*}
\mathcal{L}=\frac{1}{2}(\partial _{\mu }\varphi )(\partial ^{\mu }\varphi )+%
\frac{1}{2}(\partial _{\mu }\chi )(\partial ^{\mu }\chi )-\frac{\lambda }{4!}%
\varphi ^{4}-\frac{\lambda ^{\prime }}{4!}\chi ^{4}-\frac{\lambda ^{\prime
\prime }}{4}\varphi ^{2}\chi ^{2}
\end{equation*}%
and endow the scalar $\varphi $ with the usual prescription and $\chi $ with
the prescription (\ref{newdistr}). So doing, $\chi $ does not contribute to
the cuts, nor to the initial and final states. However, it does contribute
to the loop diagrams. If we integrate $\chi $ out, we obtain an effective
renormalizable, unitary, nonlocal theory of the self-interacting scalar
field $\varphi $.

\section{Quantum gravity with a dimensionless gauge coupling}

\setcounter{equation}{0}\label{qg}

In this section, we consider the theory described by the Lagrangian%
\begin{equation}
-2\kappa ^{2}\frac{\mathcal{L}_{\text{QG}}}{\sqrt{-g}}=2\Lambda _{C}+\zeta R-%
\frac{\gamma }{M^{2}}R_{\mu \nu }R^{\mu \nu }+\frac{1}{2M^{2}}(\gamma -\eta
)R^{2},  \label{lqg}
\end{equation}%
with $\zeta >0$ and $\gamma <0$. Although the form of $\mathcal{L}_{\text{QG}%
}$ coincides with the one of the well-known theory of refs. \cite{stelle},
we want to quantize it in a new way.

For the time being, we neglect the cosmological constant. At $\eta =0$ the
graviton propagator reads%
\begin{equation}
\langle h_{\mu \nu }(p)\hspace{0.01in}h_{\rho \sigma }(-p)\rangle _{\eta
=0}^{\text{free}}=\frac{iM^{2}}{2p^{2}(\zeta M^{2}+\gamma p^{2})}(\eta _{\mu
\rho }\eta _{\nu \sigma }+\eta _{\mu \sigma }\eta _{\nu \rho }-\eta _{\mu
\nu }\eta _{\rho \sigma }).  \label{propaf}
\end{equation}%
To have perturbative unitarity in this case, we proceed as explained in the
previous section, which means that we convert (\ref{propaf}) into%
\begin{equation*}
\left\{ \frac{1}{p^{2}+i\epsilon }-\frac{\gamma (\zeta M^{2}+\gamma p^{2})}{%
[(\zeta M^{2}+\gamma p^{2})^{2}+\mathcal{E}^{4}]_{\text{LW}}}\right\} \frac{i%
}{2\zeta }(\eta _{\mu \rho }\eta _{\nu \sigma }+\eta _{\mu \sigma }\eta
_{\nu \rho }-\eta _{\mu \nu }\eta _{\rho \sigma }),
\end{equation*}%
where $\mathcal{E}$ is the fake LW scale, which must tend to zero pretty
much like the width $\epsilon $. Similarly, at $\eta \neq 0$ we find%
\begin{equation}
\langle h_{\mu \nu }(p)\hspace{0.01in}h_{\rho \sigma }(-p)\rangle _{\eta }^{%
\text{free}}=\langle h_{\mu \nu }(p)\hspace{0.01in}h_{\rho \sigma
}(-p)\rangle _{\eta =0}^{\text{free}}-\frac{i\eta M^{2}}{2(p^{2})^{2}}\frac{%
(p^{2}\eta _{\mu \nu }+2p_{\mu }p_{\nu })(p^{2}\eta _{\rho \sigma }+2p_{\rho
}p_{\sigma })}{(\zeta M^{2}+\gamma p^{2})[\zeta M^{2}+(\gamma -3\eta )p^{2}]}%
.  \label{propag}
\end{equation}%
Assuming that $\gamma <3\eta $, the extra contribution can be cured by
turning (\ref{propag}) into%
\begin{equation*}
-\frac{i\eta }{2}\frac{(p^{2}\eta _{\mu \nu }+2p_{\mu }p_{\nu })(p^{2}\eta
_{\rho \sigma }+2p_{\rho }p_{\sigma })}{[(p^{2})^{2}+\mathcal{E}^{4}]_{\text{%
LW}}}\frac{M^{2}(\zeta M^{2}+\gamma p^{2})(\zeta M^{2}+(\gamma -3\eta )p^{2})%
}{[(\zeta M^{2}+\gamma p^{2})^{2}+\mathcal{E}^{4}]_{\text{LW}}[(\zeta
M^{2}+(\gamma -3\eta )p^{2})^{2}+\mathcal{E}^{4}]_{\text{LW}}}.
\end{equation*}

A similar procedure must be applied to the propagators of the Faddeev-Popov
ghosts.

The theory (\ref{lqg}) is the unique renormalizable higher-derivative theory
of quantum gravity whose gauge coupling $\tilde{\kappa}=\kappa M$ is
dimensionless with respect to the high-energy power counting. Indeed, if we
define $\tilde{h}_{\mu \nu }=h_{\mu \nu }/M$ and expand the metric tensor as 
$g_{\mu \nu }=\eta _{\mu \nu }+2\tilde{\kappa}\tilde{h}_{\mu \nu }$, the
Lagrangian $\mathcal{L}_{\text{QG}}$ behaves like $\sim (\square \tilde{h}%
)^{2}$ at high energies, times dimensionless constants. With the
prescriptions just given, the theory is also perturbatively unitary up to
corrections due to the cosmological constant.

We know that in general the cosmological constant is turned on by the
radiative corrections, which prevents us from proving perturbative unitarity
in a strict sense. Modified models with an identically vanishing
cosmological constant might exist. For example, it is likely possible to
build supersymmetric extensions of the theory (\ref{lqg}) that have one-loop
exact beta functions or are even finite, because similar constructions are
familiar in supersymmetric theories of fields of spins 0, 1/2 and 1 \cite%
{west}. If such models are finite, the cosmological constant can be switched
off at no cost. If they have one-loop exact beta functions, extra conditions
have to be imposed in order to fulfill the renormalization group chain (\ref%
{chain}) that follows from $\Lambda _{C}=0$. Probably, the conditions are
simple enough to admit nontrivial solutions, as we found in sections \ref{QG}
and \ref{matt} for the superrenormalizable theory (\ref{LQG}) and its
coupling to matter. In all such cases, we may provide examples of theories
quantum gravity (coupled to matter) with a dimensionless gauge coupling and
an identically vanishing cosmological constant, where the proof of unitarity
can be carried out to the very end. However, it is unlikely that the
ultimate theory of nature will have an identically vanishing cosmological
constant, so we must be prepared to accept that there may be a small
unitarity anomaly in the universe.

\section{Conclusions}

\label{concl}

\setcounter{equation}{0}

In this paper we have studied the main options for a consistent, local
quantum field theory of the gravitational interactions. Superrenormalizable
higher-derivatives theories of gravity can be built as Lee-Wick models and
formulated as nonanalytically Wick rotated Euclidean theories. They are
perturbatively unitary when the cosmological constant vanishes. The simplest
example is encoded in formula (\ref{LQG}), provided the parameters satisfy
suitable restrictions. The other models of this class can be build by adding
more higher-derivatives and fulfilling the Lee-Wick unitarity conditions.
The possibilities are infinitely many, which raises the question of
uniqueness.

A better possibility is the theory (\ref{lqg}), because it has a
dimensionless gauge constant, which makes it unique and more similar to the
gauge theories that describe the other interactions of nature. The
Lagrangian (\ref{lqg}) cannot be quantized in the conventional Lee-Wick way.
This lead us to introduce a new concept, the fake degrees of freedom, and a
prescription different from the usual one. Taking advantage of the Lee-Wick
idea, a ghost (or a normal degree of freedom) can be turned into a fake
degree of freedom, which does not contribute to the physical spectrum and
does not propagate through the cuts of the cutting equations. So doing, the
ghosts of higher-derivative gravity can be eliminated.

The renormalization of the theory (\ref{lqg}) is obviously richer than the
one of a superrenormalizable theory like (\ref{LQG}), because nontrivial
radiative corrections to the beta functions and the anomalous dimensions are
expected to all orders. Again, this is similar to what we know from the
other gauge theories that successfully describe nature.

If we accept that the gauge couplings are dimensionless, then there is only
one theory of the four interactions of nature, made of the gauge sector of
the standard model coupled to the quantum gravity theory (\ref{lqg}).
Constraining the matter sector is obviously harder.

In several models it is possible to turn the cosmological constant off to
all energies, consistently with the renormalization group. However, the more
realistic models have a nonvanishing cosmological constant, where the proof
of unitarity cannot be carried out to the very end in a strict sense. The
cosmological constant might be the anomaly of unitarity and this might be
the reason why it is so small in the universe.

It can be interesting to study the phenomenological implications of the
theory (\ref{lqg}) coupled to the standard model. Some arguments existing in
the literature (see for example ref. \cite{strumia}) might survive after
switching to the correct formulation of the theory, others might have to be
reconsidered.

\vskip 12truept \noindent {\large \textbf{Acknowledgments}}

\vskip 2truept

We are grateful to U. Aglietti and M. Piva for useful discussions.

\end{document}